# Measurement of Resistance Standards by a Precision LCR Meter at Frequencies up to 2 MHz

Jürgen Schurr[1] and Shakil A. Awan[2]

[1] Physikalisch-Technische Bundesanstalt (PTB), Bundesallee 100, 38116 Braunschweig, Germany
[2] Nanomaterials and Devices Laboratory, School of Engineering, Computing and Mathematics, University of Plymouth, Devon PL4 8AA, United Kingdom

E-mail: Juergen.Schurr@ptb.de and shakil.awan@plymouth.ac.uk

*Abstract*—We report on two-terminal-pair and four-terminal-pair test measurements of a 10-kΩ resistance standard by means of a commercial precision inductance-capacitance-resistance (LCR) meter at frequencies up to 2 MHz. In the case of a two-terminal-pair configuration, we demonstrate that the effect of 2.5 m long measuring cables, which are inevitable for some special applications, can be corrected in the whole frequency range up to 2 MHz with an uncertainty of $9 \times 10^{-6}$ ($k = 1$) relative to the dc resistance value. Furthermore, the systematic effects of the LCR meter were investigated. Though distinctly larger than the resolution of the LCR meter, these effects are accurately reproducible. As such it should be possible in the future to calibrate the LCR meter against a well-known calculable high-frequency resistance standard with an uncertainty close to the few-parts-per-million type-A uncertainty of the LCR meter.

## I. Introduction

Recent years have seen a rising interest in precision impedance measurements and calibrations at frequencies up to 2 MHz or even 100 MHz [1]–[12], and this leads accordingly to an increasing demand for international intercomparisons [13]. The measurements can be carried out with commercial instruments, such as precision LCR meters and Impedance Analyzers [1]–[7]. These instruments have excellent features and make access to the higher frequency range easier than with manual high-frequency (HF) bridges developed by NMIs [4]–[10]. Even though the demanded uncertainties increase with frequency, the measurements are still challenging. Pioneering work with Impedance Analyzers at frequencies up to 110 MHz was published in [10]–[12] but apart from [1] and [7], little research has been done with precision LCR meters in the frequency range up to 2 MHz.

To tap the full potential of a precision LCR meter and to reduce its uncertainty, it is necessary to develop a better understanding of its inherent capabilities and systematic effects. This is the motivation behind this work and represents a key step on the path to traceable LCR-meter measurements, which will in turn enable new impedance metrology applications.

In this study, we present measurements made by a commercial precision LCR meter (Keysight E4980A[*]) at frequencies up to 2 MHz. Its relative uncertainty as quoted by the manufacturer is a combination of its base uncertainty of $5 \times 10^{-4}$ and additional contributions that depend on the measuring range, the measuring voltage, the frequency, and the length of the measuring cables. However, the reproducibility of this instrument over a period of a few weeks is much better than its systematic uncertainty. For resistances on the order of 10 kΩ, an LCR meter precisely calibrated using a reliable HF calculable resistance standard puts an uncertainty of a few parts per million within reach over the entire frequency range up to 2 MHz. The average user of an LCR meter is well advised to use the manufacturer´s specifications, but those willing to invest more effort can get much more out of this instrument.

This paper is an extension of a Summary Paper presented at the conference CPEM 2024 [14] and is organized as follows: In Section II, the basic properties of the precision LCR meter used in this work are discussed. Section III and Section IV then look into the cable effect of two-terminal-pair (2TP) and four-terminal-pair (4TP) impedance standards. Section V is dedicated to the measurement of the cable parameters, while Section VI presents measurements of a conventional 10 kΩ resistance standard and investigates the systematic effects of the LCR mete.

## II. The precision LCR meter

The Keysight E4980A[*] LCR meter is a four-terminal-pair instrument that uses four Bayonet Nut Connectors (BNC) and applies an ac voltage to the device under measurement and calculates a complex impedance as the ratio of the ac voltage measured at the potential-high port and the ac current at the current-low port (measured as a voltage drop across a resistor). To meet the 4TP defining conditions, internal ammeter circuits automatically balance the current at the potential ports to zero. (The input impedance of the ammeter circuits is virtually zero, so they do not affect the measurement.) The result can be displayed either as a complex impedance value or converted to the type of impedance that is closest to the model of the device under measurement, for example, series or parallel combinations of resistance, inductance, or capacitance. A simplified schematic circuit diagram is shown in Fig. 1. Further technical details on the instrument can be found in [15].

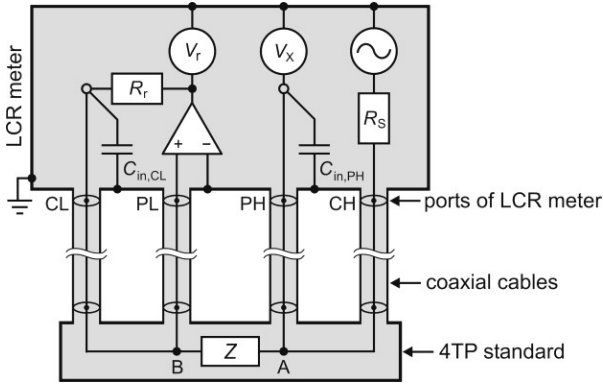

Fig. 1. Simplified circuit diagram of the precision LCR meter with a 4TP impedance standard. $R_r$ is a range resistor, $R_S$ a source resistor, and $C_{in,CL}$ and $C_{in,PH}$ are internal parasitic capacitances. A and B are the star points inside the 4TP standard. Two open circles represent the defining points inside the LCR meter.

The frequency range of the LCR meter is from 20 Hz to 2 MHz. The frequency can be set either manually to a single value or automatically to a sequence of selectable values (a so-called frequency sweep). The resolution of the LCR meter measuring a resistance of 10.0 kΩ is 0.01 Ω at a single frequency but 0.1 Ω at a frequency sweep as used in this work (corresponding to 10 parts per million).

The effect of long measuring cables on an impedance measurement is highly dependent on frequency (as will be shown below). Therefore, it is necessary to consider the relative accuracy of the frequency of the internal generator, which is specified by the manufacturer to be $1 \times 10^{-4}$. However, the frequency was measured under laboratory conditions and after the warm-up period of the LCR meter by a frequency counter referenced to an atomic clock. In this scenario, it showed a relative deviation from nominal of only $4 \times 10^{-6}$ over the entire frequency range, which is typical for a good quartz oscillator and which satisfies our most demanding requirements. Since this cannot be guaranteed for every instrument and every ambient condition, we advise other users to exercise care and to check the frequency before use.

The LCR meter can apply an open, short and load correction [15]. In the case of a test fixture (or a 4TP-1TP adapter) attached to the LCR meter, the open and short measurement determines the residual impedance and the stray capacitance inside that fixture and corrects the subsequent impedance measurements. This is a useful feature, but it must not be confused with a cable correction. The reference plane remains at the internal defining points of the LCR meter. This is why we use the open and short correction only in special cases where it is appropriate. The load correction requires that a precisely known reference standard be temporarily connected to the LCR meter. For each frequency, the LCR meter calculates a transfer function from the entered reference value and the actual measurement of that standard. This means that the "load correction" is a calibration that includes a possible load of the internal source by the impedance under measurement. As yet, we have not used this function.

In addition, the software menu offers the option of applying a calculated 4TP cable correction, though this option is limited to standard cables of a few standard lengths [15]. We do not use this function because (i) we need special (miniature and cryogenic) cables of non-standard length, (ii) the parameters of individual cables of the same type may vary due to production tolerances, and (iii) connectors and feedthroughs are not considered by the software, so the calculated cable correction would not be sufficiently accurate for our need.

Finally, the data measured by the LCR meter is read out by a computer via an opto-isolated data bus (to avoid a ground loop which may cause excess noise and fault currents).

### III. Effect of cables on a 4TP resistance standard

To calculate the effect of coaxial cables on the measurement of a 4TP resistance standard (Fig. 1), we refer to a transmission line model for incoming and reflected electromagnetic waves [4], [5], [16]. The 4TP impedance measured with long cables, $Z_{4TP}$, differs from the impedance of the component between the defining star points inside the case, $Z$, according to

$$Z_{4TP}/Z = 1/\{\cosh(X_{PH}) + Y_{in,PH}(Z_{PH}/Y_{PH})^{1/2}\sinh(X_{PH})\}$$
$$\times 1/\{\cosh(X_{CL}) + Y_{in,CL}(Z_{CL}/Y_{CL})^{1/2}\sinh(X_{CL})\}$$

with $X_i = (Z_i Y_i)^{1/2}, Z_i = (R_i + j\omega L_i), Y_i = \omega C_i(j + \tan\delta_i)$

and $Y_{in,i} = \omega C_{in,i}(j + \tan\delta_{in,i})$ each for i = PH and CL. (1)

The hyperbolic functions in (1) describe the voltage drop along the cables due to the incoming and the reflected waves. $Z_i$ and $Y_i$ are the impedance and the admittance of the defining potential-high (PH) and current-low (CL) cables, respectively. $R_i$ is the series resistance, $L_i$ the series inductance, and $\tan\delta_i$ the dissipation factor of the respective defining cable. $C_{in,i}$ are the internal capacitances of the LCR meter indicated in Fig. 1. A brief derivation of (1) is given in the Appendix. Usually, the dissipation factor of both the defining cables and the internal capacitances are small enough to be neglected, which simplifies the subsequent analysis.

The argument $X_i$ of the hyperbolic functions in (1) is equal to the product of the respective cable length and the propagation constant of the cable which is approximately equal to $(2\pi)/\lambda$ with $\lambda$ being the wavelength. In our case, $\lambda/(2\pi)$ amounts to about 16 m at 2 MHz and is much larger than the longest cables used in this work, i.e., $X_i \ll 1$. We therefore expand the hyperbolic functions into a Taylor series and consider only the first two terms:

$$Z/Z_{4TP} =$$
$$\{1 + Z_{PH}Y_{PH}/2 + [Z_{PH}Y_{PH}]^2/24 + Z_{PH}Y_{in,PH}(1 + Z_{PH}Y_{PH}/6)\}$$
$$\times \{1 + Z_{CL}Y_{CL}/2 + [Z_{CL}Y_{CL}]^2/24 + Z_{CL}Y_{in,CL}(1 + Z_{CL}Y_{CL}/6)\}$$
(2)

The first-order terms $Z_iY_i/2$ are well known in the low-frequency regime [6], whereas in our case we also need the second-order terms $[Z_iY_i]^2/24$ and the terms with the internal admittances. It is worth mentioning that even the first-order terms are a consequence of the finite wavelength or the finite speed of light.

Even in the case of long cables, the cable resistances in (2) do not significantly contribute to the real part of the impedance under measurement, so the cable correction can be simplified. The real part of the impedance $Z$, denoted by $R$, is then calculated from the measured 4TP resistance, $R_{4TP}$, according to

$$R/R_{4TP} = \{1 - \omega^2 L_{PH}C_{PH}/2 + [\omega^2 L_{PH}C_{PH}]^2/24 \\ - \omega^2 L_{PH}C_{in,PH}(1 - \omega^2 L_{PH}C_{PH}/6)\} \\ \times \{1 - \omega^2 L_{CL}C_{CL}/2 + [\omega^2 L_{CL}C_{CL}]^2/24 \\ - \omega^2 L_{CL}C_{in,CL}(1 - \omega^2 L_{CL}C_{CL}/6)\} \quad (3)$$

This is the final equation for a 4TP measurement of a resistance standard.

The cable parameters in (3) include not only the external measuring cables but also the internal leads from the front panel of the LCR meter to the internal defining points, which are not known to a normal user. Making use of the fact that the cable correction (3) depends non-linearly on the cable length, the internal lead lengths and the internal capacitances can be determined as follows: A 4TP impedance standard is connected to the LCR meter with short cables while one defining cable is successively extended and the impedance standard is measured at each cable length. The numerical value of the internal capacitance and the internal lead length are free parameters which are manually adjusted in such a way that the cable-corrected 4TP impedance becomes independent of the external cable length. Then, the same procedure is repeated for the other defining cable. We find a value of the internal capacitances of the order of 60 pF at the PH port and 130 pF at the CL port. Furthermore, it has been verified that the non-defining PL and CH cable do not contribute to the cable correction (as expected according to (2)).

## IV. Effect of Cables on a 2TP Resistance Standard

To enable 2TP measurements, the 4TP ports of the LCR meter must be adapted to a 2TP configuration. To keep the 4TP cable correction of the adapter as small as possible, a standard BNC T-connector is connected to each defining port of the LCR meter and connected to the associated non-defining port by a short cable, as shown in Fig. 2. The resulting 4TP cable correction of the adapter can be calculated by (3) and is dominated by the small inductance of the defining adapter ports combined with the internal capacitances of the LCR meter. The correction amounts to about $-130 \times 10^{-6}$ at 2 MHz and is considered small but significant. Consequently, the internal capacitances of the LCR meter have to be determined with a relative uncertainty of about 4%, which is easily feasible.

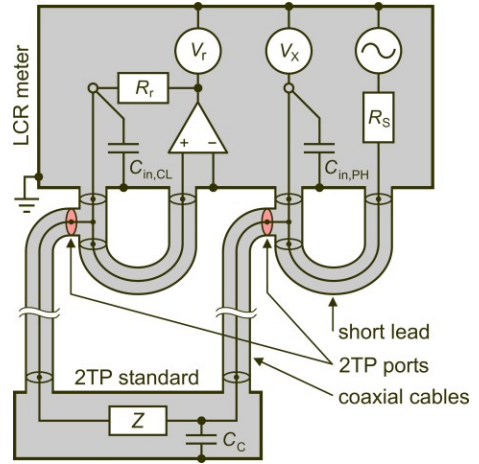

Fig. 2. The precision LCR meter configured for a measurement of a 2TP impedance standard with long coaxial cables. $C_C$ is the parasitic case capacitance at the high side of the impedance standard.

In addition to the adapter, a 2TP cable correction from the adapter ports to the 2TP impedance standard is required, which can be calculated analogously to the 4TP case. Using the same approximations as above, we get

$$R/R_{2TP} = \\ 1/\{1 - \omega^2 L_H C_H/2 - \omega^2 L_H C_C + R_H/R_0 + [\omega^2 L_H C_H]^2/24\} \\ \times 1/\{1 - \omega^2 L_L C_L/2 + R_L/R_0 + [\omega^2 L_L C_L]^2/24\}. \quad (4)$$

The indices H and L refer to the high and low side. In contrast to (3), the cable correction terms now appear in the denominator and include terms with the case capacitance at the high side, $C_C$, as well as terms with the cable resistance in ratio to the resistance of the standard. Note that the internal capacitances of the LCR meter do not contribute to the 2TP cable correction.

## V. Measurement of Cable Parameters

The cable correction of both 2TP and 4TP impedance standards requires precise knowledge of the parameters of the defining cables. These can be measured by the same LCR meter using a 4TP-1TP adapter as shown in Fig. 3. Each defining cable is sequentially connected to the 1TP port with its other end being either open-circuited to measure the cable capacitance and the associated dissipation factor or short-circuited to measure the cable resistance and the cable inductance. Of course, the impedance and admittance of the short- and open-circuited 1TP adapter port has to be subtracted. It is also important to consider all connectors involved. Since the cable parameters are frequency-dependent, the cable parameters must be measured at the same set of frequencies as the impedance standard.

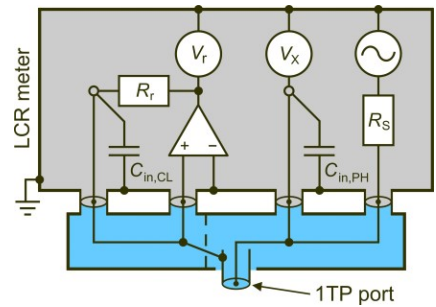

Fig. 3. The precision LCR meter configured for a measurement of cable parameters by a 4TP-1TP adapter (shown in light blue). The dashed line indicates a shield between the high and the low side.

To keep the 4TP cable correction of the adapter as small as possible, we use a special adapter which can be directly connected to the LCR meter as shown in Fig. 3. For long cables, the adapter correction is small but significant (about $-2 \times 10^{-4}$ at 2 MHz). However, the 4TP adapter correction of the measured cable admittance and cable impedance corresponds to a multiplication and a division, respectively, by the same term. Because the cable correction of a 4TP measurement (2) includes only products of a cable admittance and a cable impedance, the effect of the 4TP-1TP adapter cancels exactly. In the case of the cable correction of a 2TP measurement (4), a small contribution of the 4TP-1TP adapter remains but is fully negligible even for cables with a length of a few meters. This greatly simplifies the analysis of the data and the uncertainty.

The measured impedance of the short-circuited cables, $Z_{short}$, and the measured admittance of the open-circuited cables, $Y_{open}$, (after subtraction of the impedance and admittance of the 4TP-1TP-adapter) are affected by the propagation time of the incoming and reflected waves along the cable under measurement. This effect can be calculated from the transmission line model [16] according to

$$Z_{short,i} = (Z_i/Y_i)^{1/2} \tanh([Z_i Y_i]^{1/2}) \qquad (5)$$

$$Y_{open,i} = (Y_i/Z_i)^{1/2} \tanh([Z_i Y_i]^{1/2}) \qquad (6)$$

with the index i indicating the particular defining cable. In our case, the approximation $Z_i Y_i \ll 1$ is applicable. Therefore, we expand the tanh function into a Taylor series and consider only the first two terms. This gives

$$Z_{short,i} = Z_i \left(1 - \frac{1}{3} Z_i Y_i + \frac{2}{15} [Z_i Y_i]^2\right) \qquad (7)$$

$$Y_{open,i} = Y_i \left(1 - \frac{1}{3} Z_i Y_i + \frac{2}{15} [Z_i Y_i]^2\right). \qquad (8)$$

Using $Z_i = R_i + j\omega L_i$ and $Y_i = \omega C_i(j + \tan\delta_i)$ and considering only relevant terms, we get

$$R_{short,i} = R_i \left(1 + \frac{2}{3} \omega^2 L_i C_i + \frac{2}{15} [\omega^2 L_i C_i]^2\right) \qquad (9)$$

$$L_{short,i} = L_i \left(1 + \frac{1}{3} \omega^2 L_i C_i + \frac{2}{15} [\omega^2 L_i C_i]^2\right) \qquad (10)$$

$$C_{open,i} = C_i \left(1 + \frac{1}{3} \omega^2 L_i C_i + \frac{2}{15} [\omega^2 L_i C_i]^2\right) \qquad (11)$$

$$\tan\delta_{open,i} = \tan\delta_i + \frac{1}{3} \omega R_i C_i. \qquad (12)$$

To get an approximate solution of this set of coupled quadratic equations, we define $\Delta_i = L_{short,i} C_{open,i}$ and $L_i C_i = \Delta_i (1 + x)$. A

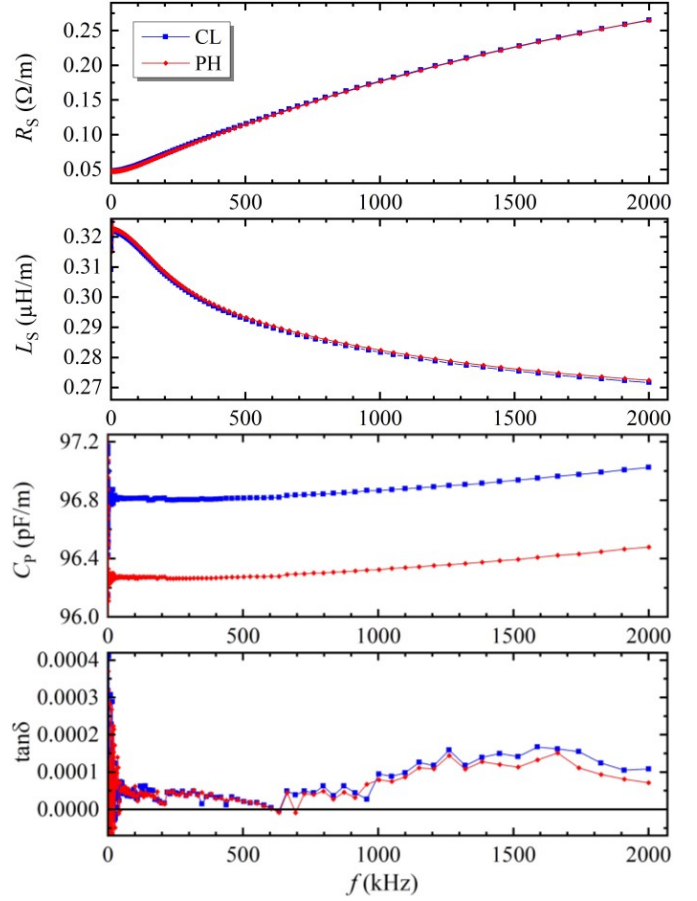

Fig. 4. The corrected cable parameters per meter of 2.5 m long coaxial cables (type Arnoflex 58 LSNH 50 ohms*) as a function of frequency.

Taylor expansion of the product of (10) and (11) for small $x$ and neglecting terms with a power of the frequency higher than four gives $x \approx -2\omega^2\Delta_i/3$. Substituting this for the $L_i C_i$ terms at the right-hand side of (9) to (12), we get the final equations for the corrected cable parameters with respect to the 1TP port of the 4TP-1TP adapter:

$$R_i = R_{short,i} / \left(1 + \frac{2}{3} \omega^2 \Delta_i - \frac{14}{45} [\omega^2 \Delta_i]^2\right) \qquad (13)$$

$$L_i = L_{short,i} / \left(1 + \frac{1}{3} \omega^2 \Delta_i - \frac{4}{45} [\omega^2 \Delta_i]^2\right) \qquad (14)$$

$$C_i = C_{open,i} / \left(1 + \frac{1}{3} \omega^2 \Delta_i - \frac{4}{45} [\omega^2 \Delta_i]^2\right) \qquad (15)$$

$$\tan\delta_i = \tan\delta_{open,i} - \frac{1}{3} \omega R_{short,i} C_{open,i}. \qquad (16)$$

Since the dissipation factor is usually small and does not significantly contribute to the cable correction, it is justified to neglect higher-order terms in (16). The terms $\omega^2 \Delta_i$ in (13) to (15) are equal to $(\omega l/c)^2$ with $l$ the cable length and $c$ the speed of light. This shows that the corrections are due to the finite propagation time of the electromagnetic waves along the cable under measurement.

For the 2.5 m long coaxial cables used in this work, the relative corrections of the cable parameters according to (13) to (15) are in the range of (1 to 2)% at 2 MHz, which is substantial but still manageable. The final results of the cable parameters are shown in Fig. 4. The frequency dependence is due to the skin effect and in reasonable agreement with a numerical HF calculation. Not all types of coaxial cables are equally suitable for HF applications, so the characterization of a particular cable (including the shield tightness) is a useful quality test.

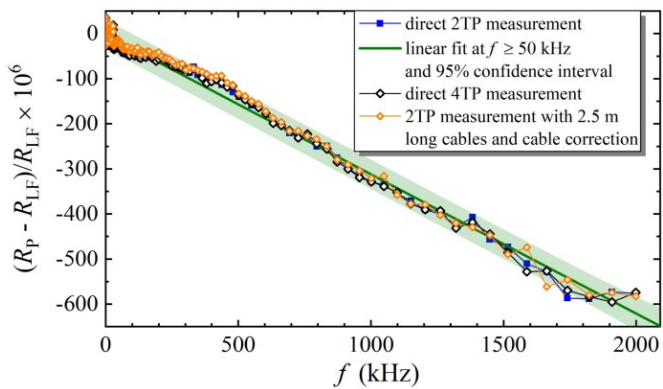

Fig. 5. The 10-kΩ resistance standard measured in different configurations as a function of frequency with respect to the respective low-frequency value, $R_{LF}$. The LCR meter was set to a long measuring time and averaged 10 measurements at each frequency. The voltage applied was 0.2 V (rms).

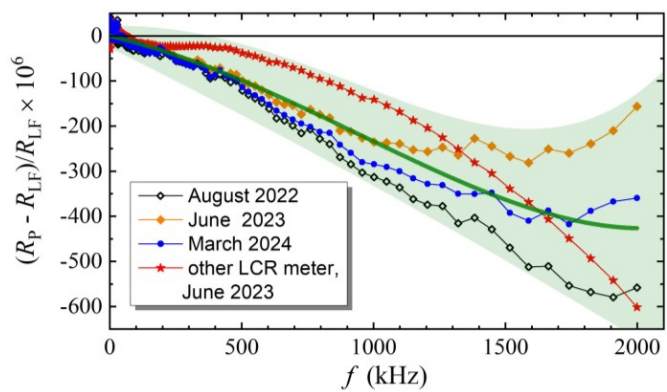

Fig. 7. Direct 4TP measurements of the 10-kΩ resistance standard as a function of frequency with respect to the respective low-frequency value, $R_{LF}$, at different dates as indicated. The black data points are from Fig. 5. The red data points were taken with another LCR meter of the same model and in the same configuration. The thick green line is a 4$^{th}$ order polynomial fit to all data at frequencies above 100 kHz. The light green band is an estimate of a 95% confidence interval as a guide to the eye.

## VI. Experimental Results

A commercial 10 kΩ 4TP resistance standard based on a single SMD resistor (Keysight 42038A*) was used as a test object. This standard is made in such a way that it can be directly connected to the four ports of the LCR meter. It is also possible to carry out 2TP measurements by using a 4TP-2TP adapter. In addition, 4TP and 2TP measurements were carried out with 2.5 m long coaxial cables. Then the parameters of the defining cables were measured at the same set of frequencies as for the resistance standard and the measurements were corrected as described in Section III to V.

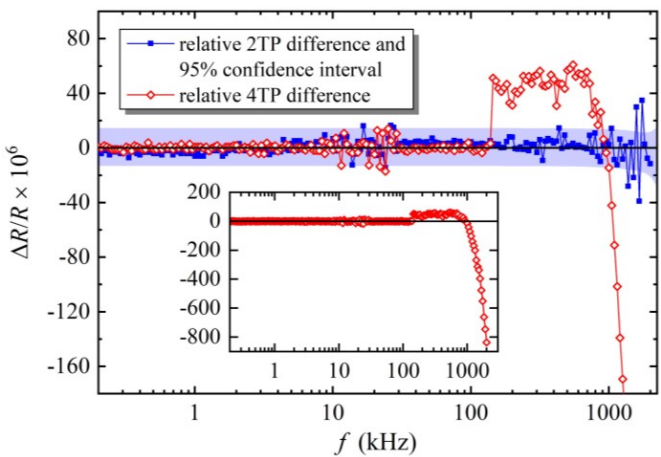

Fig. 6. The relative difference of the 10-kΩ resistance standard when measured with 2.5 m long cables and cable correction and when measured directly, in both cases as a 2TP and a 4TP configuration and as a function of frequency. The inset shows the 4TP data on a larger scale of the ordinate.

The results in Fig. 5 show that both the direct 2TP and 4TP measurements and the 2TP measurement with 2.5 m long cables and a cable correction agree with each other over the entire frequency range within a standard deviation of $4.4 \times 10^{-6}$ (at coverage factor $k = 1$). Fig. 6 shows the residuals of the 2TP and the 4TP measurement when measured with 2.5 m long coaxial cables and cable correction and when measured directly. The 2TP results demonstrate that the effect of 2.5 m long cables can be corrected in the whole frequency range up to 2 MHz with an excellent uncertainty of about $9 \times 10^{-6}$ ($k = 1$) relative to the dc resistance value. In contrast, the 4TP measurements with the same defining cables show discrepancies: First, the data shows a discontinuity of $46 \times 10^{-6}$ at that frequency at which the measured resistance exceeds the 10 kΩ range so that the LCR meter applies an automatic range change. Second, the data shows a deviation above 700 kHz that strongly increases with frequency and cable length but which is absent in a direct 4TP measurement (see Fig. 5). As the most plausible explanation, we attribute this effect to an imperfectly implemented zero-current condition at the potential ports of the LCR meter (see Fig. 1). Note that the quantum Hall resistance in multiple-series connection scheme [17] is a 2TP resistance whereas nearly all imaginable 4TP applications can be done with 1 m long measuring cables. Then, the corresponding systematic error would be reduced to $20 \times 10^{-6}$ at 2 MHz and a correction with a reasonable uncertainty could be applied.

Whereas the measurements are accurately reproducible over a period of a few weeks, substantial changes occur over a period of several months as shown in Fig. 7. (We point out that when the LCR meter was not used, it was always in stand-by mode so that the internal thermostat was continuously in operation.) The long-term changes are therefore an unavoidable property of the LCR meter. The average magnitude of the systematic variation increases from $7 \times 10^{-6}$ at 10 kHz to $86 \times 10^{-6}$ at frequencies up to 1.2 MHz and $246 \times 10^{-6}$ at 2 MHz (each relative to the dc resistance value and at $k = 1$). Within this uncertainty, the frequency dependence of the 10-kΩ standard is linear, which can be explained by dissipative parasitic capacitances [18]. The absence of a large quadratic frequency dependence indicates that the series inductance of the resistance standard is small, which is plausible because the standard is based on a thin-film SMD resistor without a fine-structured meander as common at metal-foil resistors [18].

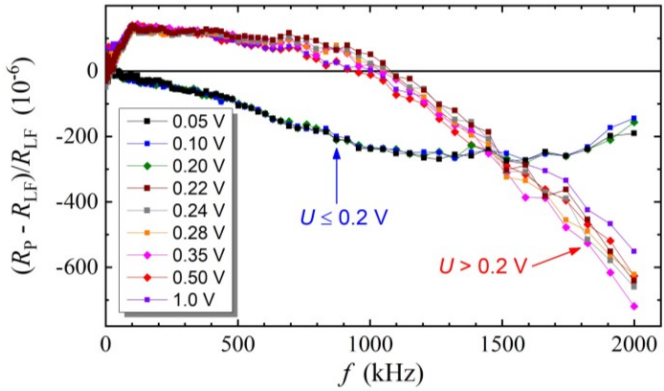

Fig. 8. Direct 4TP measurements of the 10-kΩ resistance standard as a function of frequency at successively increased voltages as indicated. The data were measured at one run within a few hours.

Furthermore, when the measuring voltage is successively increased from a low value (for example, from 50 mV to 1 V), the measured frequency dependence of the 10-kΩ resistance standard changes abruptly above a threshold of 0.2 V by up to a few hundred parts per million, as shown in Fig. 8. The reason is that at this threshold, the current-low detector automatically switches another multiplier resistor to an internal operational amplifier to keep the signal in the optimum range [15]. Each multiplier resistor substantially changes the accuracy with which the defining conditions are met and, in addition, may have a different frequency dependence. At voltages above 0.2 V, the non-physical artifacts are clearly larger while the signal-to-noise ratio is no better. Even at a low voltage of 50 mV, the signal-to-noise ratio is still good (Fig. 8). The measurements for this work were therefore carried out at 0.2 V. To prevent unperceived range changes due to noise, the manufacturer built the LCR meter such that adjacent ranges of the LCR meter slightly overlap and range changes are made with a small hysteresis [15]. Therefore, before the voltage is set to 0.2 V, we always set it temporarily to a much smaller value.

Finally, the effective parallel capacitance $C_P$ of the resistance standard was measured. The results of the direct 4TP and 2TP measurements are shown in Fig. 9. At frequencies below 10 kHz, the measured parallel capacitance shows an apparent divergence. This is a systematic error that occurs at low frequencies because the reactance $1/(\omega C_P)$ in parallel to the resistance becomes larger than the phase-angle accuracy of the LCR meter. An additional artifact occurs at a frequency of 1 kHz due to an automatic range change. (The other LCR meter of the same model mentioned above shows both effects with a similar magnitude but opposite sign, which also confirms that these effects are not true features of the standard.) To still get a reliable value of the parallel capacitance at a frequency around 1 kHz, the frequency dependence of $C_P$ in the trustworthy range above 200 kHz can be extrapolated to 1 kHz with an appropriate uncertainty. This method assumes that the true frequency dependence of $C_P$ above 1 kHz is small. This assumption is plausible because according to the Kramer-Kronig relation, the frequency dependence of $C_P$ is proportional to the associated dissipation factor, which for common SMD resistors is small. (Alternatively, the systematic error of the LCR meter in the frequency range below 20 kHz could be determined by an impedance simulator [19] so that a correction could be applied.) In the case of 2TP and 4TP measurements with 2.5 m long cables, the measured and cable-corrected parallel capacitance is found to agree with the direct measurement within about 5 fF ($k = 1$), which is excellent.

This method to determining the effective parallel capacitance of a 10-kΩ resistance standard has an estimated uncertainty of 15 fF ($k = 1$) and is consistent with PTB's calibration capabilities (with an uncertainty of 58 fF at $k = 1$). The dominating uncertainty contribution is the phase-angle inaccuracy of the LCR meter which is determined from a measurement of low-loss capacitance standards. When this effect would be corrected, the uncertainty would be 7 fF ($k = 1$), corresponding to an uncertainty of the time constant of 0.09 ns, which as far as we know, is smaller than for all other measurement and calculation methods (for example [20]–[22]). However, before this method can be used for the legal calibration of time constants, a deeper evaluation is required.

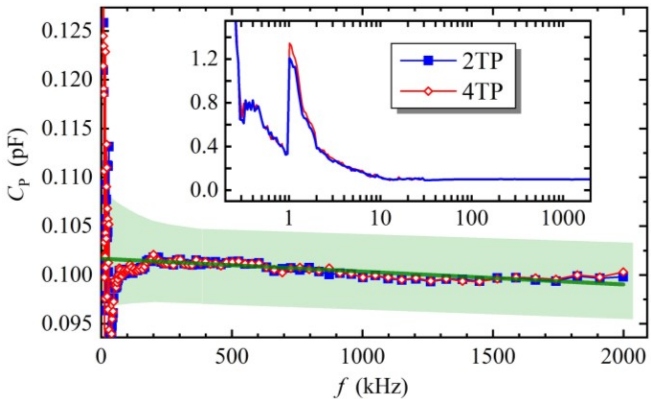

Fig. 9. The parallel capacitance $C_P$ of the 10-kΩ resistance standard measured by the LCR meter as a function of frequency. The green line is a linear least-square fit to all data above 200 kHz and the light green band is an estimate of a confidence interval as a guide to the eye. The inset shows the same data on a logarithmic frequency scale.

## VII. Conclusion

Using a 10-kΩ resistance standard, we investigated the systematic effects of a commercial precision LCR meter. These effects are smaller than the uncertainty specified by the manufacturer but much larger than the resolution of the instrument. The main point here is that all these effects are accurately reproducible (at least in the short-term) and should thus be correctable in the future through calibration against a high-frequency calculable resistance standard. (Such a standard is currently under development by the authors.)

Furthermore, we demonstrated that the effect of 2.5 m long measuring cables on a 2TP resistance can be accurately corrected. Such long cables are inevitable for future 2TP precision measurements of the quantum Hall resistance of GaAs and graphene devices [23]–[25] at frequencies up to 2 MHz. Such measurements would represent a world premiere. The hope is that the frequency dependence of the quantum Hall resistance will turn out to be zero within the resolution of the LCR meter. This would demonstrate that the quantum Hall resistance is a universal quantum standard of impedance. Furthermore, this will enable new applications for impedance metrology, such as graphene [26], 2D materials, and biosensing [27], as well as precise calibration of conventional high-frequency impedance standards.

Appendix

In this appendix, (1) is briefly derived. The voltage and the current measured at the internal ports of the LCR meter (see figure 1) are named $V_{PH}$, $I_{PH}$, $V_{CL}$, and $I_{CL}$. They are related to the voltage and the current at the star points A and B inside the impedance standard named $V_A$, $I_A$, $V_B$, and $I_B$. This relation is determined by the transmission lines connecting the internal star points of the impedance standard to the defining points inside the LCR meter and by the internal admittances of the LCR meter, $Y_{in,PH}$ and $Y_{in,CL}$. We assume that the transmission lines and the internal admittances at the potential-high and the current-low side can be considered as linear and passive two-port networks, and that the transmission lines are homogenous. Then, we can use the matrix and vector formalism [1], [2], [16], [28] and express the relation between the corresponding quantities at the star points inside the impedance standard and the defining points inside the LCR meter by a matrix of a transmission line, $M_1$, and a matrix of the internal admittance, $M_2$, each for the potential-high and the current-low side:

$$\begin{bmatrix} V_{PH} \\ I_{PH} \end{bmatrix} = M_{2PH} \cdot M_{1PH} \cdot \begin{bmatrix} V_A \\ I_A \end{bmatrix} \text{ and } \begin{bmatrix} V_{CL} \\ I_{CL} \end{bmatrix} = M_{2CL} \cdot M_{1CL} \cdot \begin{bmatrix} V_B \\ I_B \end{bmatrix}$$

with $M_{2i} = \begin{bmatrix} 1 & 0 \\ Y_{in,i} & 1 \end{bmatrix}$, $M_{1i} = \begin{bmatrix} A_i & B_i \\ C_i & D_i \end{bmatrix}$,

$A_i = D_i = \cosh(\sqrt{Z_i Y_i})$, $B_i = \sqrt{Z_i/Y_i} \sinh(\sqrt{Z_i Y_i})$

and $C_i = \sqrt{Y_i/Z_i} \sinh(\sqrt{Z_i Y_i})$ each for i = PH and CL.

The impedance between the internal star points A and B is defined as $Z = V_A/I_B$. The 4TP impedance measured by the LCR meter is given by $Z_{4TP} = V_{PH}/I_{CL}$ with the auxiliary conditions $I_{PL} = 0$, $I_{PH} = 0$, and $V_{PL} = 0$. The latter also means $V_B = 0$. From this and taking into account that $A_i D_i - B_i C_i = 1$, we get

$$\frac{Z_{4TP}}{Z} = \frac{V_{PH} I_B}{I_{CL} V_A} = \frac{1}{(B_{PH} Y_{in,PH} + D_{PH}) \cdot (B_{CL} Y_{in,CL} + D_{CL})}$$

Inserting the definition of the matrix components $B_i$ and $D_i$ given above, we finally get (1).

Acknowledgment

The authors would like to thank F. Overney and M. Agustoni from METAS for support of (1) and the analysis described in the Appendix. They would also like to thank R. Judaschke for support on the transmission line model and for numerical model calculations of coaxial cables, O. Kieler for providing another LCR meter of the same model, and N. Abraham for valuable technical support.

\* The identification and characterization of specific commercial instruments and components in this article does not imply recommendation or endorsement by PTB or the University of Plymouth, nor does it imply that these instruments and components are necessarily the best available for the given purpose.